\documentclass[aps,prl,twocolumn,superscriptaddress,longbibliography]{revtex4-1}

\usepackage{graphicx}
\usepackage{dcolumn}
\usepackage{bm}
\usepackage{amssymb}
\usepackage{microtype}
\usepackage{xfrac}
\usepackage{array}

\newcommand{\ket}[1]{\ensuremath{|#1\rangle}}

\usepackage{mathtools}
\DeclarePairedDelimiter{\evdel}{\langle}{\rangle}
\newcommand{\ev}{\evdel}

\begin{document}

\title{Magneto-elastic induced vibronic bound state in the spin ice pyrochlore Ho$_2$Ti$_2$O$_7$} 

\author{J.~Gaudet}
\affiliation{Department of Physics and Astronomy, McMaster University, Hamilton, ON, L8S 4M1, Canada}

\author{A.~M.~Hallas}
\affiliation{Department of Physics and Astronomy, McMaster University, Hamilton, ON, L8S 4M1, Canada}

\author{C.~R.~C.~Buhariwalla}
\affiliation{Department of Physics and Astronomy, McMaster University, Hamilton, ON, L8S 4M1, Canada}

\author{G.~Sala}
\affiliation{Department of Physics and Astronomy, McMaster University, Hamilton, ON, L8S 4M1, Canada}
\affiliation{Neutron Scattering Division, Oak Ridge National Laboratory, Oak Ridge, Tennessee 37831, USA}

\author{M.~B.~Stone}
\affiliation{Neutron Scattering Division, Oak Ridge National Laboratory, Oak Ridge, Tennessee 37831, USA}

\author{M.~Tachibana}
\affiliation{National Institute for Materials Science, 1-1 Namiki, Tsukuba 305-0044, Ibaraki, Japan}

\author{K.~Baroudi}
\affiliation{Department of Chemistry, Princeton University, New Jersey 08544, USA}

\author{R.~J.~Cava}
\affiliation{Department of Chemistry, Princeton University, New Jersey 08544, USA}
\affiliation{Princeton Materials Institute, Princeton University, New Jersey 08544, USA}

\author{B.~D.~Gaulin}
\affiliation{Department of Physics and Astronomy, McMaster University, Hamilton, ON, L8S 4M1, Canada}
\affiliation{Canadian Institute for Advanced Research, Toronto M5G 1M1, Canada}
\affiliation{Brockhouse Institute for Materials Research, Hamilton, ON L8S 4M1 Canada}

\date{\today}

\begin{abstract}
The single ion physics of Ho$_2$Ti$_2$O$_7$ is well-understood to produce strong Ising anisotropy, which is an essential ingredient to its low-temperature spin ice state. We present inelastic neutron scattering measurements on Ho$_2$Ti$_2$O$_7$ that reveal a clear inconsistency with its established single ion Hamiltonian. Specifically, we show that a crystal field doublet near 60~meV is split by approximately 3~meV. Furthermore, this crystal field splitting is not isolated to Ho$_2$Ti$_2$O$_7$ but can also be found in its chemical pressure analogs, Ho$_2$Ge$_2$O$_7$ and Ho$_2$Sn$_2$O$_7$. We demonstrate that the origin of this effect is a vibronic bound state, resulting from the entanglement of a phonon and crystal field excitation. We derive the microscopic Hamiltonian that describes the magneto-elastic coupling and provides a quantitative description of the inelastic neutron spectra. 
\end{abstract}

\maketitle
The pyrochlore oxide Ho$_2$Ti$_2$O$_7$ is a quintessential dipolar spin ice material~\cite{bramwell2001spin}. In this system, the  Ho$^{3+}$ moments sit on the vertices of a corner-sharing tetrahedral network. Each Ho$^{3+}$ moment possesses strong local Ising anisotropy such that each spin is constrained to point either towards (``in'') or away from (``out'') the center of their respective tetrahedra~\cite{harris1997geometrical,Rosenkranz2000}. At low temperatures, below $\theta_{CW} = 2$~K, the Ho$^{3+}$ moments adopt a two-in/two-out arrangement, a spin structure that exactly maps onto the proton configuration in water ice~\cite{pauling1935structure,harris1997geometrical}. In addition to moments with local Ising anisotropy decorating the pyrochlore lattice, the other key ingredient for the dipolar spin ice state is an effective ferromagnetic coupling between the Ho$^{3+}$ moments, which originates from long-range dipolar interactions~\cite{den2000dipolar}. A decade after the initial discovery of the dipolar spin ice state in Ho$_2$Ti$_2$O$_7$~\cite{harris1997geometrical}, a host of studies revealed that the elementary spin excitations in this state are emergent magnetic monopoles~\cite{Castelnovo2008,jaubert2009signature,Fennel2009,morris2009dirac,ladak2010direct,giblin2011creation}.   

Although Ising anisotropy is a key ingredient to the spin ice state, it has been the subject of relatively few investigations~\cite{Rosenkranz2000,jana2000crystal,RuminyCEF}. Naturally, most studies on Ho$_2$Ti$_2$O$_7$ have focused on the collective spin behavior within the ice state. Here, taking advantage of recent advances in the instrumentation of time-of-flight neutron spectroscopy, we take a closer and more comprehensive look at the crystal electric field (CEF) scheme of Ho$_2$Ti$_2$O$_7$ and two of its sister spin ice materials, Ho$_2$Sn$_2$O$_7$~\cite{matsuhira2000low,kadowaki2002neutron} and Ho$_2$Ge$_2$O$_7$~\cite{zhou2012chemical,hallas2012statics}. In doing so, we observe a splitting of a high energy CEF excitation, a feature that could not be observed in previous neutron scattering works due to their lower energy resolution. We show that this split excitation cannot be accounted for by either a pure CEF excitation or by a pure phonon excitation.  We conclude that its origin is a magneto-elastic coupling induced vibronic bound state, a hybridized excitation resulting from the entanglement of a phonon and a crystal field excitation. 

The vibronic bound state is a quantum phenomena that is challenging to unambiguously identify because it requires a detailed independent knowledge of a material's phonon dispersion and CEF scheme~\cite{loewenhaupt2003coupling}. Consequently, there have been few definitive examples of materials with this phenomenology and these have generally been limited to materials with uncomplicated CEF spectra. For example, the landmark observation of a vibronic bound state in CeAl$_2$~\cite{Thalmeier1982,Thalmeier1984} was aided by the fact that only a single excited CEF level is expected in this material. Another notable example is the high $T_C$ superconductor NdBa$_2$Cu$_3$O$_{7-\delta}$, where isotopic substitution of oxygen was used to shift the phonon energy, thereby altering the nature of the bound state~\cite{heyen1991coupling}. Our work expands the notion of vibronic bound states to the pyrochlore magnet Ho$_2$Ti$_2$O$_7$, a material with a considerably more complex CEF scheme. We propose that in addition to being prototypical spin ice materials, Ho$_2$Ti$_2$O$_7$ and its sister holmium pyrochlores are also quintessential examples of materials that exhibit a vibronic bound state.

\begin{figure*}[tbp]
\linespread{1}
\par
\includegraphics[width=7in,height=4in]{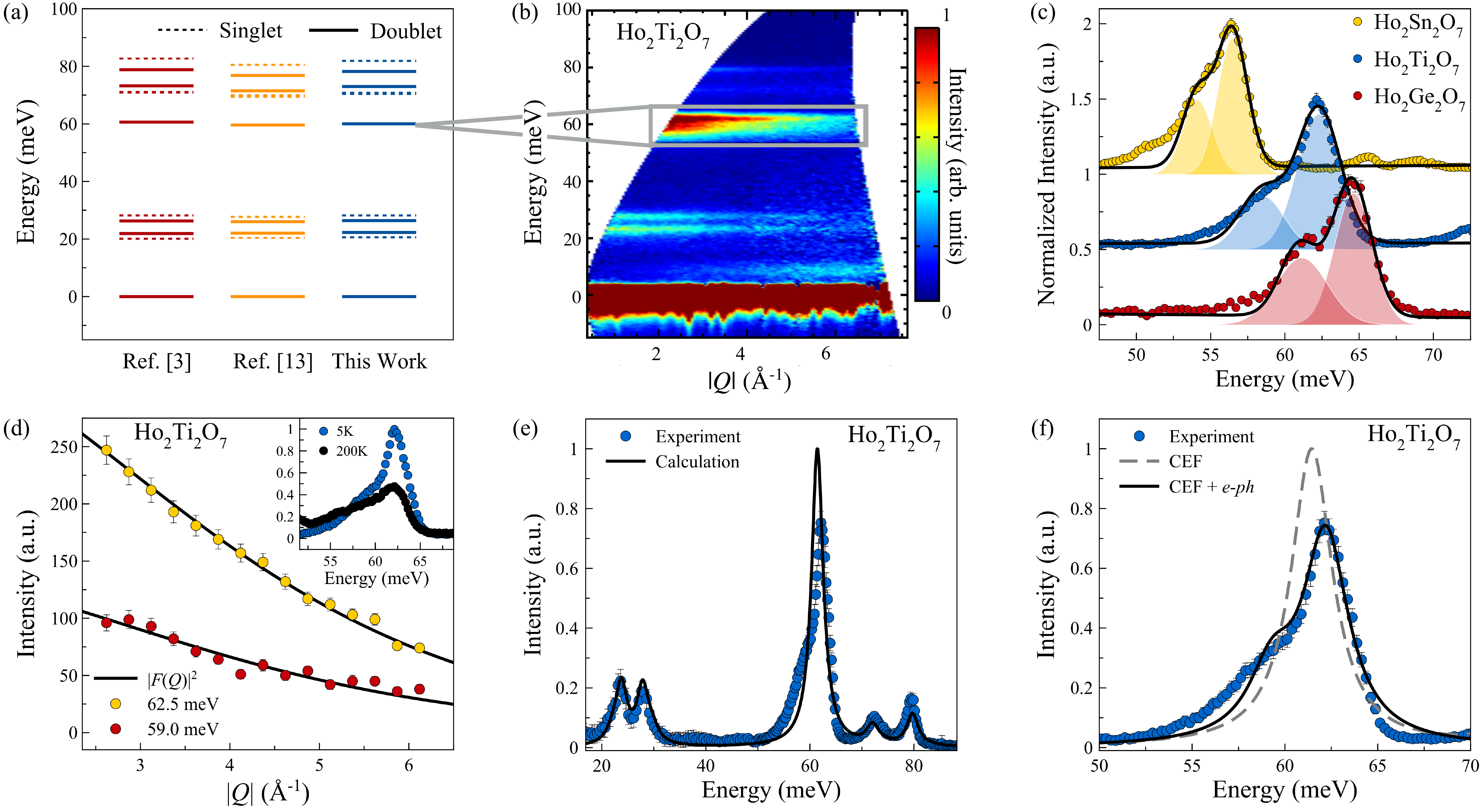}
\par
\caption{(a) The CEF energy scheme for Ho$_2$Ti$_2$O$_7$ deduced by Rosenkranz \emph{et al.}~\cite{Rosenkranz2000}, Ruminy \emph{et al.}~\cite{RuminyCEF} and our current work. This scheme was determined via inelastic neutron scattering measurements, such as the one shown in (b), a spectrum that was collected with $E_i=150$~meV at $T=5$~K. (c) Integrated scattering from 3 to 4~\AA$^{-1}$, reveals a splitting of the CEF level near 60 meV in each of Ho$_2$Ge$_2$O$_7$, Ho$_2$Ti$_2$O$_7$, and Ho$_2$Sn$_2$O$_7$. (d) Both of the excitations near 60~meV follow the magnetic form factor for Ho$^{3+}$, where the intensities were extracted by fitting to the sum of two Gaussians, as depicted in (c). The temperature dependence is also consistent with a magnetic origin, as shown in the inset. (e) Our refined CEF Hamiltonian for Ho$_2$Ti$_2$O$_7$ gives a good description of the experimental data, but does not capture the splitting near 60~meV. (f) The splitting can be modeled by including a magneto-elastic term in our Hamiltonian, which describes the hybridization of the CEF level with a phonon.}
\label{Refinement}
\end{figure*}

In the holmium pyrochlores, the local spin anisotropy depends only on the composition of the CEF ground state, which in turn depends on the symmetry and strength of the CEF at the Ho$^{3+}$ site. In the case of $D_{3d}$ point group symmetry, which is appropriate for the rare earth site in the pyrochlore lattice, the CEF Hamiltonian can be expressed as follows~\cite{Hutchings,Prather,Freeman1962,Walter1984}:
\begin{eqnarray}
\mathcal{H}_{CEF}= A^0_2\alpha_J\ev{r^2}\hat{O}^0_2 + A^0_4\beta_J\ev{r^4}\hat{O}^0_4 +  
\nonumber
\\
A^3_4\beta_J\ev{r^4}\hat{O}^3_4 + A^0_6\gamma_J\ev{r^6}\hat{O}^0_6 +
\nonumber
\\
A^3_6\gamma_J\ev{r^6}\hat{O}^3_6 + A^6_6\gamma_J\ev{r^6}\hat{O}^6_6,
\label{eq: HCEF}
\end{eqnarray}
where the $\hat{O}^n_m$ are Stevens operators~\cite{Stevens}. The $\alpha_J$, $\beta_J$ and $\gamma_J$ are reduced matrix elements calculated in Ref.~\cite{Stevens}. The terms $\ev{r^n}$ are the expected value of the distance between the nucleus and the 4$f$ electron shell taken to the $n$th power, and are tabulated in Ref.~\cite{freeman1979dirac}. According to the symmetry of the CEF Hamiltonian, the $2J+1=17$ states of the spin-orbit ground state are split into 6 doublets and 5 singlets. The energy configuration and composition of these CEF states are controlled by the CEF parameters $A^m_n$, which can be estimated from point-charge calculations~\cite{Hutchings} or experimentally determined by inelastic neutron spectroscopy~\cite{bertin2012crystal}. 

The single ion properties of Ho$_2$Ti$_2$O$_7$ have been previously studied with inelastic neutron scattering by Rosenkranz \emph{et al.}~\cite{Rosenkranz2000} and Ruminy \emph{et al.}~\cite{RuminyCEF}. In Fig. 1(a), we reproduce the CEF energy scheme of Ho$_2$Ti$_2$O$_7$ based on these earlier works. In both cases, the ground state is a well-isolated non-Kramers doublet. The first set of excited crystal field levels, which are located between 20 and 30 meV, is made up of two singlets and two doublets, followed by an isolated doublet at 60~meV and, finally, two doublets and three singlets between 70 and 90~meV. The four closely spaced doublet/singlet pairs are the products of a distorted cubic oxygen environment, where in a precisely cubic environment, these would form four triplets~\cite{lea1962}. The most important result for the low temperature magnetism is that the ground state doublet consists of two states with dominant $\ket{m_J=\pm8}$, resulting in a strong local Ising anisotropy.

We have carried out inelastic neutron scattering measurements to investigate the CEF schemes of three holmium pyrochlores: Ho$_2$Ge$_2$O$_7$, Ho$_2$Ti$_2$O$_7$ and Ho$_2$Sn$_2$O$_7$. In each of these three materials the magnetism originates from Ho$^{3+}$ and the variation from non-magnetic Ge to Ti to Sn primarily modifies the cubic lattice parameter. Ho$_2$Ge$_2$O$_7$ and Ho$_2$Sn$_2$O$_7$ can be considered as positive and negative chemical pressure analogs of Ho$_2$Ti$_2$O$_7$, respectively. The inelastic neutron scattering measurements were performed using a next generation chopper spectrometer, SEQUOIA, at the Spallation Neutron Source at Oak Ridge National Laboratory~\cite{granroth2010sequoia}. The full inelastic spectra for each of these three materials can be found in the Supplemental Material as well as further experimental details.  

We begin by considering the $E_i = 150$~meV spectra for Ho$_2$Ti$_2$O$_7$, which is shown in Fig.~1(b). This data set was collected at $T =5$~K, which is well below the threshold for thermally populating excited crystal field states. Thus, at this temperature, all CEF excitations must originate from the ground state. A handful of CEF excitations (at approximately 22, 26, 60, 71 and 78~meV) can be immediately identified based on their lack of dispersion and intensity that decreases as a function of $|\mathbf{Q}|$. By comparing Fig.~1(a) and Fig.~1(b), we see that the energies of these excitations are in good agreement with the predicted energy schemes of the previous CEF studies on Ho$_2$Ti$_2$O$_7$~\cite{Rosenkranz2000,RuminyCEF}. However, a closer examination of the scattering spectra near 60~meV reveals a striking inconsistency: there are two distinct excitations near 60~meV in our new experimental data instead of the single well isolated CEF excitation predicted by the Hamiltonians of both Rosenkranz \emph{et al.}~\cite{Rosenkranz2000} and Ruminy \emph{et al.}~\cite{RuminyCEF}. This can be better appreciated by performing an integration over the energy interval in question, which is shown in Fig. 1(c). A similar structure is observed for the equivalent excitations in Ho$_2$Ge$_2$O$_7$ and Ho$_2$Sn$_2$O$_7$, which occur at respectively slightly higher and lower energy transfers. In each case, the lower energy excitation has less than half the intensity of the higher energy one. In the paragraphs that follow, we will address the plausible origins for these two excitations and ultimately show that their origin is a magneto-elastic coupling induced vibronic bound state.

We will first demonstrate that a pure phonon excitation is not the origin of either of the excitations near 60~meV. This can be verified by analyzing the $|\mathbf{Q}|$ dependence of their scattered intensity, which is presented in Fig.~1(d) for Ho$_2$Ti$_2$O$_7$. The intensities of the two excitations were extracted by taking integrations of the data set shown in Fig.~1(b) in 0.2~\AA$^{-1}$ intervals, which were then fit to the sum of two Gaussian functions. Examples of such fits for each of the three holmium pyrochlores is given in Fig.~1(c). In all three samples, the intensities of both excitations are observed to decrease monotonically with $|\mathbf{Q}|$, a behavior that is unmistakably inconsistent with the $\mathbf{Q}^2$ dependence expected for a pure phonon excitation~\cite{Squires}. Instead, as shown in Fig.~1(d), the $|\mathbf{Q}|$ dependence of both excitations agrees well with the magnetic form factor for Ho$^{3+}$, which is the expected behavior for a CEF excitation. Furthermore, as seen in the inset of Fig.~1(d), the intensities of both excitations decrease when the temperature is raised from 5 to 200 K, a temperature dependence that is also inconsistent with the behavior of a phonon excitation.

Given that the 60 meV excitations have intensities that follow the magnetic form factor for Ho$^{3+}$, we next considered the possibility that they are both pure CEF excitations. Thus, we attempted to refine a new CEF Hamiltonian for Ho$_2$Ti$_2$O$_7$ that includes two or more CEF levels around 60~meV. We followed the same method used in Ref.~\cite{Rosenkranz2000,gaudet2018effect}, which consists of varying the set of CEF parameters, $A^n_m$ in Eqn.~1, until a minimum value of $\chi^2$ is reached between the computed and experimental CEF spectrum. However, this search did not yield any satisfactory result; over an extensive range of parameter space all solutions with multiple transitions near 60 meV contained glaring inconsistencies with other features in the experimental spectrum. Furthermore, having two CEF transitions near 60 meV for Ho$_2$Ti$_2$O$_7$ is also inconsistent with point-charge calculations~\cite{Tomasello2015,bertin2012crystal}. Starting from a robust, highly constrained determination of the CEF Hamiltonian for the erbium pyrochlore Er$_2$Ti$_2$O$_7$, we apply a scaling procedure to approximate the CEF Hamiltonians of other titanate pyrochlores~\cite{gaudet2018effect}. Using this scaling argument, the calculated CEF scheme for Ho$_2$Ti$_2$O$_7$, aside from containing only a single isolated excitation near 60~meV, is in excellent qualitative agreement with our experimental data and the previous reports by Rosenkranz \emph{et al.}~\cite{Rosenkranz2000} and Ruminy \emph{et al.}~\cite{RuminyCEF}. Moreover, when this scaling procedure is performed in reverse, starting from a CEF Hamiltonian for Ho$_2$Ti$_2$O$_7$ that contains two excitations near 60~meV, the scaling gives results which are wholly inconsistent with the known CEF spectra for Tb$_2$Ti$_2$O$_7$~\cite{RuminyCEF,princep2015}, Er$_2$Ti$_2$O$_7$~\cite{gaudet2018effect} and Yb$_2$Ti$_2$O$_7$~\cite{gaudet2015neutron}. Thus, we conclude that the two excitations near 60~meV cannot be pure CEF excitations and must originate from the degeneracy breaking of the CEF doublet. 

\begin{table*}[]
\centering
\caption{Tabulated results of the CEF analysis for Ho$_2$Ge$_2$O$_7$, Ho$_2$Ti$_2$O$_7$ and Ho$_2$Sn$_2$O$_7$ including the CEF parameters $A^n_m$ (in meV), the Ising component of the local $g$-tensor, $g_z$, and the effective moment of the ground state doublet, $\mu_{CEF}$ (in $\mu_B$). We then give the calculated and experimental values of the splitting energy, $\Delta E$ (in meV), and intensity ratio, $I^{ratio}$, of the split CEF level at 60~meV. The final two columns give the refined magneto-elastic constant, $g_{0}$, and phonon energy, $\hbar \omega_{0}$ (in meV).}
\label{CEF_Solutions_Table}
\begin{tabular}{lp{1cm}<{\centering}p{1cm}<{\centering}p{1cm}<{\centering}p{0.8cm}<{\centering}p{1cm}<{\centering}p{1cm}<{\centering}p{1cm}<{\centering}p{1cm}<{\centering}p{1cm}<{\centering}p{1cm}<{\centering}p{1cm}<{\centering}p{1cm}<{\centering}p{1cm}<{\centering}p{1cm}<{\centering}}
\hline
\hline
\multicolumn{1}{r}{\textit{}} & $A^0_2$ & $A^0_4$ & $A^3_4$ & $A^0_6$ & $A^3_6$  & $A^6_6$ & $g_z$ & $\mu_{CEF}$ & $\Delta E_{exp}$ & $\Delta E_{calc}$ & $I^{ratio}_{exp}$ & $I^{ratio}_{calc}$ & $g_{0}$ & $\hbar \omega_{0}$\\ \hline
Ho$_2$Ge$_2$O$_7$                  & 64.9   & 27.3  & 185  & 1.05  & $-16.9$  & 24.0 & 19.4 & 9.7 & 3.1(2) & 2.8 & 2.5(1) & 2.5  & 0.04(1) & 65(1)\\
Ho$_2$Ti$_2$O$_7$                  & 50.3   & 26.1  & 185  & 1.05  & $-15.6$  & 20.0 & 19.6 & 9.8 & 3.5(2) & 3.6 & 2.5(1) & 2.4  & 0.04(1) & 61(1)\\
Ho$_2$Sn$_2$O$_7$                  & 59.7   & 22.7  & 191  & 0.93  & $-14.7$  & 19.0 & 19.6 & 9.8 & 2.4(2) & 2.6 & 2.1(1) & 2.1  & 0.04(1) & 55(1)\\ \hline
\hline
\end{tabular}
\end{table*}

Putting aside the origin of the split doublet for a moment, we will first describe the results of our conventional CEF analysis on the three holmium pyrochlores. For this analysis, we disregarded the splitting of the 60 meV doublet and assigned its energy as the average value, 60.8~meV for Ho$_2$Ti$_2$O$_7$, and assigned its intensity as the total value for both excitations. 
The CEF Hamiltonians obtained for all three holmium pyrochlores are given in Table 1. The calculated energies and intensities for 
Ho$_2$Ge$_2$O$_7$, Ho$_2$Ti$_2$O$_7$, and Ho$_2$Sn$_2$O$_7$ are tabulated in the Supplemental Material and result in $\chi^2$ values of 1.5, 1.0, and 1.1, respectively. In Fig.~1(a), we schematically present the energy scheme for Ho$_2$Ti$_2$O$_7$ based on our parameterization side-by-side with those of the previous studies~\cite{Rosenkranz2000,RuminyCEF}, showing that they are entirely consistent with one another. The computed inelastic neutron scattering spectra for Ho$_2$Ti$_2$O$_7$ at $T=5$~K is shown in Fig.~1(e), and provides an excellent description of the experimental data aside from the obvious discrepancy near 60 meV. The ground state CEF doublet is, as expected, highly anisotropic. The component of the local $g$-tensor perpendicular to the Ising axis is strictly zero. In fact, this is provided by symmetry for a non-Kramers doublet since, under time reversal, $J_z$ transforms like a magnetic dipole while $J_x,J_y$ transform like an electric quadrupole~\cite{Lee2012}. The component parallel to the local Ising axis ($g_z$) as well as the effective moment of the CEF ground state doublet are listed in Table 1. Neither of these quantities is observed to significantly vary with chemical pressure. Thus, as it relates to their spin ice physics, the holmium pyrochlores are effectively identical in terms of their CEF properties.

We return now to the degeneracy breaking of the CEF doublet near 60~meV. One possibility is that a symmetry reducing structural distortion could be responsible for lifting the degeneracy of this doublet. However, highly sensitive neutron Larmor diffraction measurements have shown that no such transition occurs in Ho$_2$Ti$_2$O$_7$ down to 0.5~K~\cite{RuminyLarmor}.  Exchange interactions could also produce a degeneracy breaking, but the energy scale of the exchange interactions in rare earth pyrochlores is of order 1~meV~\cite{Fennel2009}, whereas the splitting of the CEF doublet is observed to persist up to at least 200~K ($k_BT = 17$~meV), as shown in the inset of Fig.~1(d). 
A final possible origin for the CEF splitting is an electron-phonon or magneto-elastic coupling, and here we have finally arrived at an explanation which, as we now show, is fully compatible with our experimental results.

A magneto-elastic interaction can be modeled by introducing a linear coupling between the phonon displacements of the system and the quadrupolar operator of the Ho$^{3+}$ ion. Including the non-interacting phonon coupling, the system can be described by the following Hamiltonian~\cite{Thalmeier1982,Thalmeier1984}:
\begin{eqnarray}
\mathcal{H}_{tot} = \mathcal{H}_{CEF} + \sum_{\mu} \hbar \omega_{\mu} (\hat{a}_{\mu}^{\dagger}\hat{a}_{\mu}+\frac{1}{2}) 
- \sum_{\mu,i} g_{\mu}\hat{U}_{\mu}\hat{O}_{i}
\end{eqnarray}   
where $\hat{U}_{\mu}=(\hat{a}_{\mu}+\hat{a}_{\mu}^{\dagger})$ is the phonon displacement operator. The operators $\hat{a}_{\mu}$ and $\hat{a}_{\mu}^{\dagger}$ correspond to the annihilation and creation of a phonon with displacement $\mu$ and energy $\hbar \omega_{\mu}$. The magneto-elastic constant is given by $g_{\mu}$, while $\hat{O}_i$ corresponds to quadrupolar operators that can be written in terms of total angular momentum operators. For the $D_{3d}$ point group symmetry, there are five quadrupolar operators with three different symmetries: $\hat{O}_{1} = 3 J_{z}^{2} - J(J+1)$, $\hat{O}_{2,3} = J_{x}^{2} - J_{y}^{2},\hspace{0.2cm}J_{x}J_{y} + J_{y}J_{x}$ and $\hat{O}_{4,5} = J_{x}J_{z} + J_{z}J_{x},\hspace{0.2cm}J_{y}J_{z} + J_{z}J_{y}$. Magneto-elastic coupling between two CEF states is only allowed if the symmetry of the quadrupolar operator and the phonon eigenvector are identical~\cite{Lovesey2000}. The symmetry of the phonon displacements in Ho$_2$Ti$_2$O$_7$ were characterized in Ref.~\cite{RuminyPhonon} and thus, symmetry analysis can be used to predict which CEF states are candidates for magneto-elastic coupling. Only the CEF states at 26~meV, 60~meV and 78~meV have quadrupolar and displacement operators of the correct symmetry to couple with the ground state~\cite{Ruminy2017}. Furthermore, we find that the matrix element involving the state at 60~meV is an order of magnitude larger than the state at 26~meV and three orders of magnitude larger than the state at 78~meV. This provides a natural explanation for why we only observe a splitting of the 60~meV CEF state. 

In order to quantify the magneto-elastic coupling, we extracted the eigenvectors associated with the ground state ($v_1,v_2$) and the 60~meV excited state ($v_3,v_4$) obtained from our CEF analysis. The non-interacting phonon contribution was approximated by a single phonon of energy $\hbar \omega_{0}$. Then, the Hamiltonian shown in Eqn.~2 was diagonalized using the following states: $\ket{v_{cef},v_{ph}}=\ket{v_{1},0},~\ket{v_{2},0},~\ket{v_{1},\hbar \omega_0},~\ket{v_{2},\hbar \omega_0},~\ket{v_{3},0},~\ket{v_{4},0}$. A splitting of the 60 meV CEF level is observed whenever the magneto-elastic constant, $g_{0}$, is non-zero and the energy of the phonon, $\hbar \omega_{0}$, is close to that of the CEF level. To determine the precise values of $g_{0}$ and $\hbar \omega_{0}$, we performed a least squares refinement of the energy difference ($\Delta E$) and the intensity ratio ($I^{ratio}$) of the split CEF level. The refined values of $g_{0}$ and $\hbar \omega_{0}$ are tabulated in Tab. 1, along with a comparison of the experimental and calculated values of $\Delta E$ and $I^{ratio}$. The resulting CEF spectra for Ho$_2$Ti$_2$O$_7$ is presented in Fig.~1(f), showing good agreement with the measured data. In each of the three holmium pyrochlores, the splitting does not produce two peaks of equal intensity, but rather the higher energy excitation is roughly twice the intensity of the lower energy excitation. This originates from the fact that the phonon involved in the coupling is located at slightly higher energy than the CEF excitation. The phonon involved in this coupling corresponds to an oxygen displacement~\cite{RuminyPhonon}, which may favor magneto-elastic coupling because the oxygen ions provide the dominant contribution to the CEF. The composition of the eigenstates for the split CEF levels, which are given in the Supplemental Material, consists of a linear combination of a pure CEF excitation ($\ket{v_{3},0}$,$\ket{v_{4},0}$) and a pure phonon excitation ($\ket{v_{1},\hbar \omega_0}$,$\ket{v_{2},\hbar \omega_0}$). 

We have demonstrated the existence of strong magneto-elastic coupling in the holmium pyrochlores. This coupling is a quantum phenomena that can be termed a vibronic bound state, the coherent propagation of a bound CEF and a phonon excitation. This effect is analogous to the exciton excitation observed in semiconductors where an electron and a hole particle are bound together. Vibronic bound states are rare~\cite{loewenhaupt2003coupling} and have been most frequently observed in cerium based intermetallics, such as CeAl$_2$ and CeCuAl$_3$~\cite{Thalmeier1982,Schedler2003,Chapon2006,Adroja2012}. Within the pyrochlore family of materials, magneto-elastic coupling has been argued as relevant in the spin liquid behavior of Tb$_2$Ti$_2$O$_7$, where vibronic excitations were also observed~\cite{Fennel2014,RuminyCEF,Constable2017}. However, in the case of Tb$_2$Ti$_2$O$_7$, where the coupling occurs at 1~meV, a quantitative description has not yet been achieved due to the complex influence of exchange interactions. Magneto-elastic coupling has also been investigated in the context of spin ice, where the monopole dynamics has been shown to depend on the spin-lattice interaction~\cite{Ruminy2017,Borzi2016,orendavc2007}. Considering the body of work including our new findings, it is clear that magneto-elastic coupling is an important interaction in the rare-earth pyrochlores and that it will play a major role in the discovery of new quantum phenomena.


\begin{acknowledgments}
This research at ORNL's Spallation Neutron Source was sponsored by the Scientific User Facilities Division, Office of Basic Energy Sciences, U.S. Department of Energy. This work was supported by the Natural Sciences and Engineering Research Council of Canada. The sample growth at Princeton University was supported by the US Department of Energy, Basic Energy Sciences, through the IQM at Johns Hopkins University, grant DE-FG02-98-ER46544.
\end{acknowledgments}

\bibliography{Ho_CEF_Bib}

\end{document}


\title{SUPPLEMENTAL MATERIAL: \\
Magneto-elastic coupling induced vibronic bound state in the spin ice pyrochlore Ho$_2$Ti$_2$O$_7$}

\author{J.~Gaudet}
\affiliation{Department of Physics and Astronomy, McMaster University, Hamilton, ON, L8S 4M1, Canada}

\author{A.~M.~Hallas}
\affiliation{Department of Physics and Astronomy, McMaster University, Hamilton, ON, L8S 4M1, Canada}

\author{C.~R.~C.~Buhariwalla}
\affiliation{Department of Physics and Astronomy, McMaster University, Hamilton, ON, L8S 4M1, Canada}

\author{G.~Sala}
\affiliation{Department of Physics and Astronomy, McMaster University, Hamilton, ON, L8S 4M1, Canada}
\affiliation{Neutron Scattering Division, Oak Ridge National Laboratory, Oak Ridge, Tennessee 37831, USA}

\author{M.~B.~Stone}
\affiliation{Oak Ridge National Laboratory, Oak Ridge, Tennessee 37831, USA}

\author{M.~Tachibana}
\affiliation{National Institute for Materials Science, 1-1 Namiki, Tsukuba 305-0044, Ibaraki, Japan}

\author{Student~Cava}
\affiliation{Department of Chemistry, Princeton University, New Jersey 08544, USA}

\author{R.~J.~Cava}
\affiliation{Department of Chemistry, Princeton University, New Jersey 08544, USA}
\affiliation{Princeton Materials Institute, Princeton University, New Jersey 08544, USA}

\author{B.~D.~Gaulin}
\affiliation{Department of Physics and Astronomy, McMaster University, Hamilton, ON, L8S 4M1, Canada}
\affiliation{Canadian Institute for Advanced Research, Toronto M5G 1M1, Canada}
\affiliation{Brockhouse Institute for Materials Research, Hamilton, ON L8S 4M1 Canada}

\date{\today}

\maketitle

\section{Experimental methods}

Powder samples of Ho$_2$Ti$_2$O$_7$, Ho$_2$Sn$_2$O$_7$, and Lu$_2$Ti$_2$O$_7$ of approximately 10~g each, were prepared by conventional solid state synthesis. The 1.8~g sample of Ho$_2$Ge$_2$O$_7$ was synthesized using high pressure techniques following the same protocol as Ref.~\cite{hallas2016xy}. The samples of Ho$_2$Ti$_2$O$_7$ and Ho$_2$Sn$_2$O$_7$ were loaded in helium atmosphere between flat aluminum plates and sealed with indium. The samples of Ho$_2$Ge$_2$O$_7$ and Lu$_2$Ti$_2$O$_7$ were sealed in the same manner, but in an annular aluminum can. Inelastic neutron scattering measurements were performed on all three holmium pyrochlores as well as the non-magnetic Lu$_2$Ti$_2$O$_7$ with the time-of-flight spectrometer SEQUOIA at the Spallation Neutron Source at Oak Ridge National Laboratory~\cite{granroth2010sequoia}. The measurements for Ho$_2$Ti$_2$O$_7$, Ho$_2$Sn$_2$O$_7$ and Lu$_2$Ti$_2$O$_7$ were performed in a closed cycle refrigerator, giving a base temperature of 5~K, while Ho$_2$Ge$_2$O$_7$ was measured in an orange ILL liquid helium cryostat, giving a base temperature of 2~K. Data sets were collected for each sample with a lower incident energy (either 45~meV or 50~meV) and a higher incident energy (120~meV or 150~meV). The fine Fermi chopper was used to give the maximum energy resolution for each incident energy, approximately 1~meV for the lower energy data sets and 3~meV for the higher energy data sets at the elastic line. An empty flat plate was also measured under an identical configuration, and was subtracted from the data. The data was reduced using Mantid~\cite{Mantid} and analyzed with DAVE~\cite{Dave}.\\

\begin{figure*}[tbp]
\linespread{1}
\par
\includegraphics[width=6in]{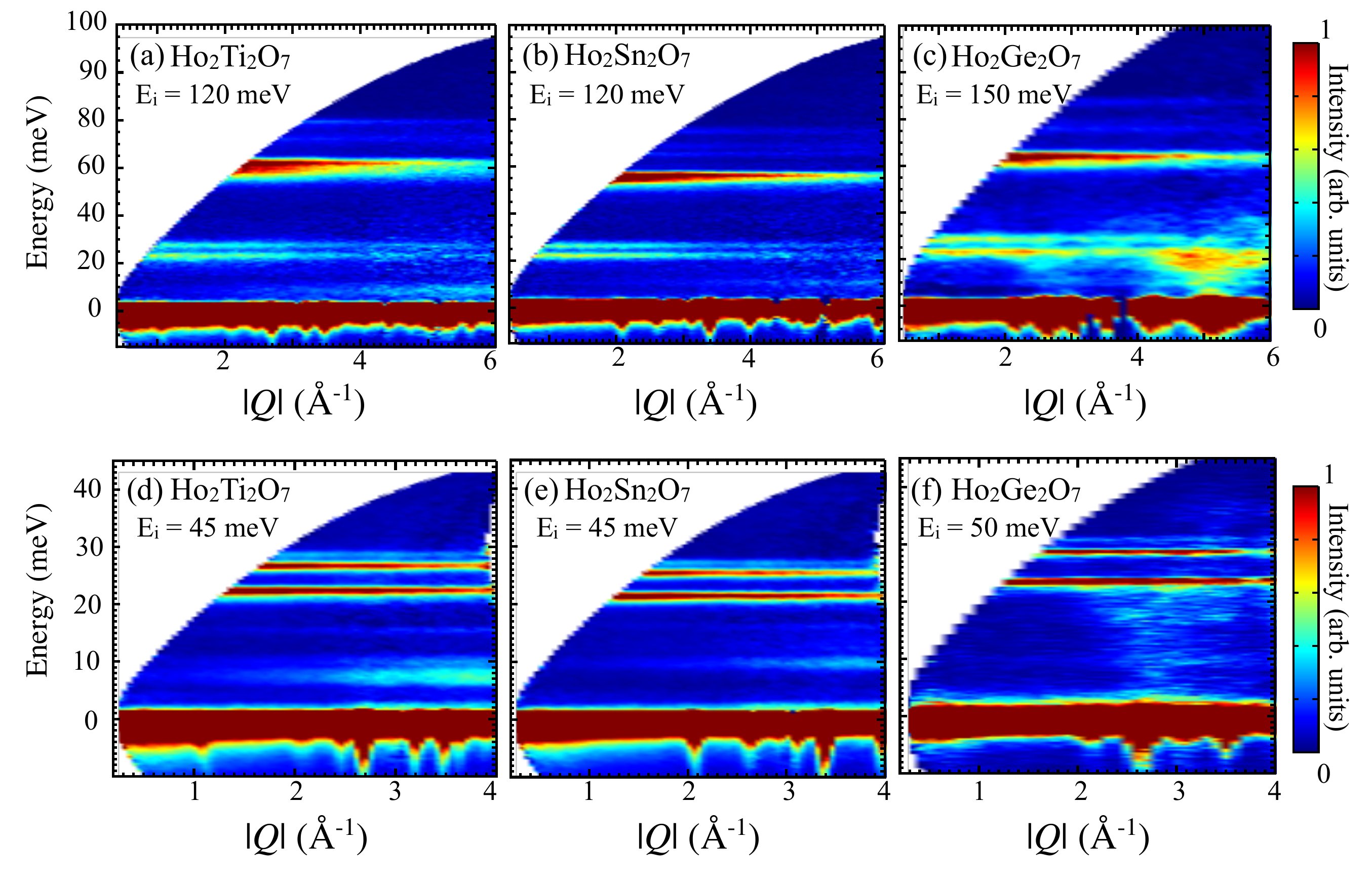}
\par
\caption{Neutron scattering spectra obtained for an incident energy of 120 or 150~meV for (a) Ho$_2$Ti$_2$O$_7$, (b) Ho$_2$Sn$_2$O$_7$ and (c) Ho$_2$Ge$_2$O$_7$ and for an incident energy of 45 or 50~meV for (e) Ho$_2$Ti$_2$O$_7$, (f) Ho$_2$Sn$_2$O$_7$ and (g) Ho$_2$Ge$_2$O$_7$. These data sets were collected at 5~K for Ho$_2$Ti$_2$O$_7$ and Ho$_2$Sn$_2$O$_7$ and 2~K for Ho$_2$Ge$_2$O$_7$.}
\label{FitBragg}
\end{figure*}

\begin{figure}[tbp]
\linespread{1}
\par
\includegraphics[width=3in]{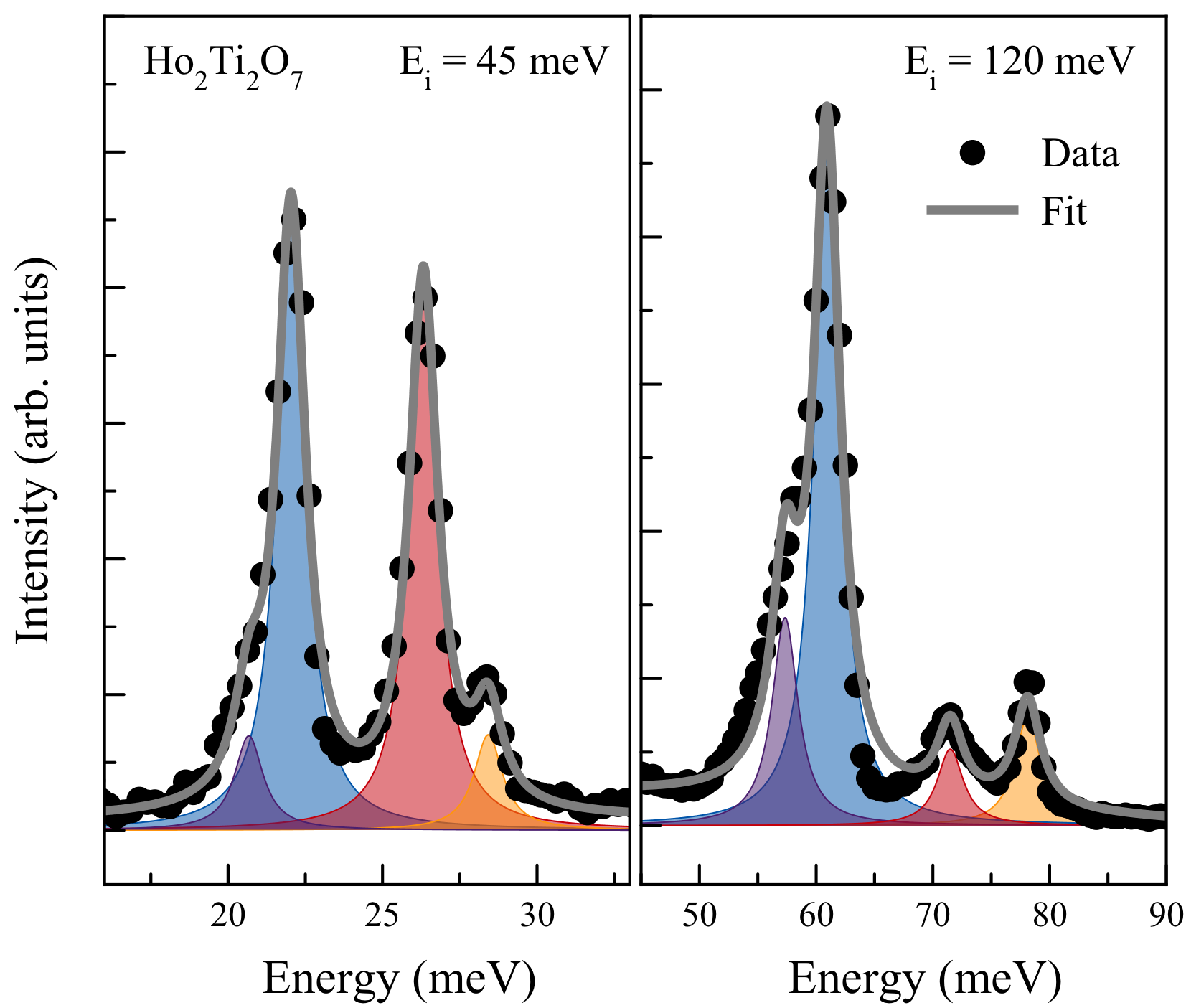}
\par
\caption{Integrated scattering spectrum for Ho$_2$Ti$_2$O$_7$ at 5~K, obtained by integrating the $E_i=45$~meV data set from 2 to 3~\AA$^{-1}$ and the $E_i=120$~meV data set from 4 to 5~\AA$^{-1}$. The energies and intensities of the CEF transitions were obtained by fitting to a series of Lorentzian functions, given by the shaded peaks, and their sum is represented by the gray line.}
\label{fig5}
\end{figure}

\section{Determination of the CEF Hamiltonians}

Inelastic neutron scattering measurements with $E_i=120$ or 150~meV (Fig.~1(a-c)) and $E_i=45$ or 50~meV (Fig.~1(d-f)) were performed in order to determine the CEF Hamiltonians of Ho$_2$Ti$_2$O$_7$, Ho$_2$Sn$_2$O$_7$ and Ho$_2$Ge$_2$O$_7$. As can be seen by comparing Fig.~1(a,d) with Fig.~1(b,e) and Fig.~1(c,f), the CEF schemes of the three holmium pyrochlores are qualitatively similar. A quantitative analysis of the CEF excitations was achieved by integrating over momentum transfer, $|\mathbf{Q}|$, between 4 and 5~\AA$^{-1}$ for $E_i=120$ or 150~meV and between 2 and 3~\AA$^{-1}$ for $E_i=45$ or 50~meV. The resultant integrated intensities are shown for Ho$_2$Ti$_2$O$_7$ in Fig.~2. Each CEF transition was fit with a Lorentzian function in order to extract its energy and scattered intensity. The resulting fit for Ho$_2$Ti$_2$O$_7$ is shown in Fig.~2. Note that the first excited CEF level is located at 20~meV ($E/k_B=230$~K) and is thus not thermally populated at 5~K. Therefore all observed CEF levels correspond to excitations out of the ground state. The CEF transition near 20~meV was clearly resolved in both the low ($E_i=45$ or 50~meV) and high ($E_i=120$ or 150~meV) energy spectra and was used to normalize the two data sets. The energy of the CEF transition near 60~meV was taken as the average energy of the split CEF excitation. The observed energies (E$_{obs}$) and scattered intensities (I$_{obs}$) fitted for all three compounds are presented in Tab.~1. The CEF Hamiltonians were refined following the same $\chi^2$ minimization method used in Ref.~\cite{gaudet2018effect}. The CEF parameters of Ho$_2$Ti$_2$O$_7$ obtained in Ref.~\cite{Rosenkranz2000} were used as the starting solution and the $\chi^2$ minimization leads to the refined CEF parameters, $A^n_m$ shown in Tab.~1 of the main manuscript. The calculated CEF transition energies and scattered intensities are given in Tab.~1.\\

\section{Phonon excitations in $Lu_2Ti_2O_7$}

The neutron scattering spectra of non-magnetic Lu$_2$Ti$_2$O$_7$ was also measured in order to characterize the phonon spectra. In particular, we expect that the phonon spectrum for Lu$_2$Ti$_2$O$_7$ should be almost identical to that of Ho$_2$Ti$_2$O$_7$, given their structural similarities and comparable cation masses. The $E_i=150$~meV spectrum of Lu$_2$Ti$_2$O$_7$ is shown in Fig.~3(a). All of the features observed in this spectrum have intensities that increase as a function of momentum transfer, $|\bold{Q}|$, characteristic of phonon excitations. We performed an integration in $\bold{Q}$ between 4 and 5~\AA$^{-1}$ and a second integration between 6 and 7~\AA$^{-1}$, the results of which are shown in Fig.~3(b). We can see that Lu$_2$Ti$_2$O$_7$ has phonon excitations centered just below 10~meV, multiple bands of phonon from 20 to 50~meV and finally, bands of phonon excitations centered around 60 and 75~meV. As mentioned in the main manuscript, the phonon spectrum of Ho$_2$Ti$_2$O$_7$ has been previously characterized using density functional theory and resonant inelastic x-ray scattering. This study showed that the low energy phonons are dominated by Ho$^{3+}$ and Ti$^{4+}$ displacements and the high energy phonons are mostly dominated by the displacements of the O$^{2-}$ ions. In particular for the magneto-elastic coupling, the phonon involved in the vibronic bound state of Ho$_2$Ti$_2$O$_7$ is centered around 60~meV and known to originate from the displacement of the O$^{2-}$ ions. It is interesting to note that in the oxide pyrochlores, the largest contribution to the CEF comes from the oxygen ions. One can speculate that phonons involving oxygen ions are more effective in coupling with a CEF excitation to induce a vibronic bound state.

\begin{figure}[tbp]
\linespread{1}
\par
\includegraphics[width=3.5in]{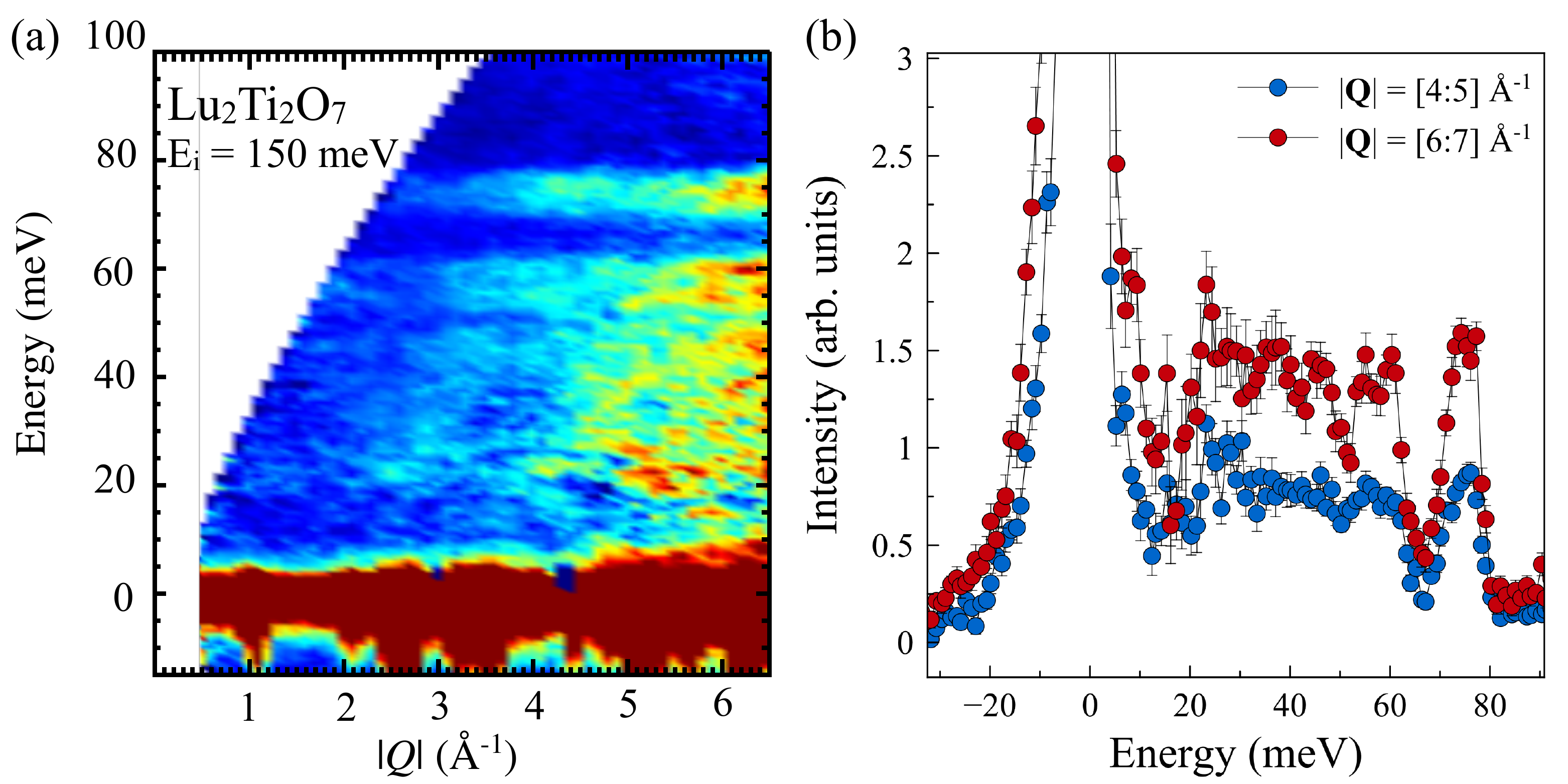}
\par
\caption{(a) Inelastic neutron scattering spectra of Lu$_2$Ti$_2$O$_7$ measured with $E_i=150$~meV at 5~K showing the phonon excitation spectrum. (b) Integrations of the scattering spectra between 4 and 5~\AA$^{-1}$ and between 6 and 7~\AA$^{-1}$ revealing several bands of phonon excitations.}
\label{fig5}
\end{figure}

\section{Diagonalization of the magneto-elastic interaction in the holmium pyrochlores}

The Hamiltonian describing the single-ion properties of the holmium ions in Ho$_2B_2$O$_7$ is given by:
\begin{eqnarray}
\mathcal{H}_{tot} = \mathcal{H}_{CEF} + \sum_{\mu} \hbar \omega_{\mu} (\hat{a}_{\mu}^{\dagger}\hat{a}_{\mu}+\frac{1}{2}) 
- \sum_{\mu,i} g_{\mu}\hat{U}_{\mu}\hat{O}_{i}
\end{eqnarray}
where the first term is the CEF contribution, the second term is the phonon contribution and the third term is the magneto-elastic coupling. The operators $\hat{a}_{\mu}$ and $\hat{a}_{\mu}^{\dagger}$ correspond to the annihilation and creation of a phonon with displacement $\mu$ and energy $\hbar \omega_{\mu}$ while $\hat{O}_i$ corresponds to quadrupolar operators (ex: $\hat{O}_{1} = 3 J_{z}^{2} - J(J+1)$). Since we are only interested in the coupling between the CEF ground state doublet ($v_1,v_2$) and the CEF excited states near 60~meV ($v_3,v_4$), we can construct a Hamiltonian that is appropriate only for the $v_1,v_2,v_3,v_4$ subspace of the CEF term. Furthermore, we assumed the magneto-elastic coupling involves only a single phonon of energy $\hbar \omega_{0}$. Within such formalism, the Hamiltonian can be simplified as: 
\begin{eqnarray}
\mathcal{H}_{tot} = \sum_{i=1,2,3,4}E_{i}\ket{v_i}\bra{v_i} + \hbar \omega_{0} \hat{a}_{0}^{\dagger}\hat{a}_{0} - \sum_{i} g_{0}\hat{U}_{0}\hat{O}_{i}
\end{eqnarray}      
where $E_{i}$ is the energy of the CEF eigenstate $v_i$. To diagonalize the magneto-elastic interaction, we use the six following eigenstates: $\ket{v_1,0}$, $\ket{v_2,0}$, $\ket{v_3,0}$, $\ket{v_4,0}$, $\ket{v_1,\hbar \omega_{0}}$ and $\ket{v_2,\hbar \omega_{0}}$ where for example, $\ket{v_1,\hbar \omega_{0}}$ describes the system with an active phonon of energy $\hbar \omega_{0}$ and the holmium ion being in its CEF ground state $v_1$. Assuming a non-zero value for $g_0$ and the phonon energy $\hbar \omega_{0}$ that is close to the energy of the CEF states $v_3$ and $v_4$, we observe a degeneracy breaking of the $v_3$ and $v_4$ CEF states. In order to determine the values of $g_0$ and $\hbar \omega_{0}$, we performed a $\chi^2$ refinement of the energy difference ($\Delta E$) and the intensity ratio ($I^{ratio}$) of the split CEF level. The refined values of $\Delta E$ and $I^{ratio}$ are given in Tab.~1 of the main manuscript. Furthermore, the new eigenstates of the system are given in Tab.~2.

\begin{table}[]
\centering
\caption{The experimentally observed and calculated values of the energies and neutron scattered intensities of the CEF levels in (a) Ho$_2$Ti$_2$O$_7$, (b) Ho$_2$Sn$_2$O$_7$ and (c) Ho$_2$Ge$_2$O$_7$. The calculated values were obtained using the refined CEF parameters found in Tab.~1 of the main manuscript.}
\label{my-label}
\begin{tabular}{ccccc}
\hline
\multicolumn{5}{|c|}{(a) Ho$_2$Ti$_2$O$_7$}                                                                                                                   \\ \hline
\multicolumn{1}{|l|}{}  & \multicolumn{1}{c|}{$\quad E_{obs} \quad$} & \multicolumn{1}{c|}{~$\enspace E_{calc} \enspace$~} & \multicolumn{1}{c|}{~$\quad I_{obs} \quad$~} & \multicolumn{1}{c|}{~$\enspace I_{calc} \enspace$~} \\ \hline
\multicolumn{1}{|c|}{$\enspace$D$\enspace$} & \multicolumn{1}{c|}{0}         & \multicolumn{1}{c|}{0}          & \multicolumn{1}{c|}{-}         & \multicolumn{1}{c|}{-}          \\ \hline
\multicolumn{1}{|c|}{S} & \multicolumn{1}{c|}{20.7(2)}   & \multicolumn{1}{c|}{20.6}       & \multicolumn{1}{c|}{0.03(2)}   & \multicolumn{1}{c|}{0.03}       \\ \hline
\multicolumn{1}{|c|}{D} & \multicolumn{1}{c|}{22.0(2)}   & \multicolumn{1}{c|}{22.3}       & \multicolumn{1}{c|}{0.19(5)}   & \multicolumn{1}{c|}{0.2}        \\ \hline
\multicolumn{1}{|c|}{D} & \multicolumn{1}{c|}{26.3(3)}   & \multicolumn{1}{c|}{26.4}       & \multicolumn{1}{c|}{0.17(5)}   & \multicolumn{1}{c|}{0.19}       \\ \hline
\multicolumn{1}{|c|}{S} & \multicolumn{1}{c|}{28.4(2)}   & \multicolumn{1}{c|}{28.2}       & \multicolumn{1}{c|}{0.03(2)}   & \multicolumn{1}{c|}{0.04}       \\ \hline
\multicolumn{1}{|c|}{D} & \multicolumn{1}{c|}{58.9(4)}   & \multicolumn{1}{c|}{59.9}       & \multicolumn{1}{c|}{1}         & \multicolumn{1}{c|}{1}          \\ \hline
\multicolumn{1}{|c|}{S} & \multicolumn{1}{c|}{71.2(4)}   & \multicolumn{1}{c|}{70.4}       & \multicolumn{1}{c|}{0.05(2)}   & \multicolumn{1}{c|}{0.04}       \\ \hline
\multicolumn{1}{|c|}{S} & \multicolumn{1}{c|}{-}         & \multicolumn{1}{c|}{70.8}       & \multicolumn{1}{c|}{0.04(2)}   & \multicolumn{1}{c|}{0.02}       \\ \hline
\multicolumn{1}{|c|}{D} & \multicolumn{1}{c|}{-}         & \multicolumn{1}{c|}{72.3}       & \multicolumn{1}{c|}{-}         & \multicolumn{1}{c|}{0.007}      \\ \hline
\multicolumn{1}{|c|}{D} & \multicolumn{1}{c|}{77.9}      & \multicolumn{1}{c|}{78.2}       & \multicolumn{1}{c|}{0.12(3)}   & \multicolumn{1}{c|}{0.11}       \\ \hline
\multicolumn{1}{|c|}{S} & \multicolumn{1}{c|}{-}         & \multicolumn{1}{c|}{81.9}       & \multicolumn{1}{c|}{-}         & \multicolumn{1}{c|}{0.03}       \\ \hline
                        & \multicolumn{1}{l}{}           & \multicolumn{1}{l}{}            & \multicolumn{1}{l}{}           & \multicolumn{1}{l}{}            \\ \hline
\multicolumn{5}{|c|}{(b) Ho$_2$Sn$_2$O$_7$}                                                                                                                   \\ \hline
\multicolumn{1}{|c|}{}  & \multicolumn{1}{c|}{$E_{obs}$} & \multicolumn{1}{c|}{$E_{calc}$} & \multicolumn{1}{c|}{$I_{obs}$} & \multicolumn{1}{c|}{$I_{calc}$} \\ \hline
\multicolumn{1}{|c|}{D} & \multicolumn{1}{c|}{0}         & \multicolumn{1}{c|}{0}          & \multicolumn{1}{c|}{-}         & \multicolumn{1}{c|}{-}          \\ \hline
\multicolumn{1}{|c|}{S} & \multicolumn{1}{c|}{20.4(2)}   & \multicolumn{1}{c|}{20.3}       & \multicolumn{1}{c|}{0.03(2)}   & \multicolumn{1}{c|}{0.03}       \\ \hline
\multicolumn{1}{|c|}{D} & \multicolumn{1}{c|}{21.3(2)}   & \multicolumn{1}{c|}{21.6}       & \multicolumn{1}{c|}{0.18(5)}   & \multicolumn{1}{c|}{0.17}       \\ \hline
\multicolumn{1}{|c|}{D} & \multicolumn{1}{c|}{25.2(2)}   & \multicolumn{1}{c|}{25.1}       & \multicolumn{1}{c|}{0.15(4)}   & \multicolumn{1}{c|}{0.17}       \\ \hline
\multicolumn{1}{|c|}{S} & \multicolumn{1}{c|}{27.0(2)}   & \multicolumn{1}{c|}{26.9}       & \multicolumn{1}{c|}{0.03(2)}   & \multicolumn{1}{c|}{0.04}       \\ \hline
\multicolumn{1}{|c|}{D} & \multicolumn{1}{c|}{53.2(4)}   & \multicolumn{1}{c|}{54.3}       & \multicolumn{1}{c|}{1}         & \multicolumn{1}{c|}{1}          \\ \hline
\multicolumn{1}{|c|}{S} & \multicolumn{1}{c|}{64.3(4)}   & \multicolumn{1}{c|}{63.8}       & \multicolumn{1}{c|}{0.07(3)}   & \multicolumn{1}{c|}{0.04}       \\ \hline
\multicolumn{1}{|c|}{S} & \multicolumn{1}{c|}{67.8(4)}   & \multicolumn{1}{c|}{67.6}       & \multicolumn{1}{c|}{0.06(3)}   & \multicolumn{1}{c|}{0.02}       \\ \hline
\multicolumn{1}{|c|}{D} & \multicolumn{1}{c|}{-}         & \multicolumn{1}{c|}{69.3}       & \multicolumn{1}{c|}{-}         & \multicolumn{1}{c|}{0.01}       \\ \hline
\multicolumn{1}{|c|}{D} & \multicolumn{1}{c|}{73.7}      & \multicolumn{1}{c|}{73.9}       & \multicolumn{1}{c|}{0.10(3)}   & \multicolumn{1}{c|}{0.06}       \\ \hline
\multicolumn{1}{|c|}{S} & \multicolumn{1}{c|}{-}         & \multicolumn{1}{c|}{76.6}       & \multicolumn{1}{c|}{-}         & \multicolumn{1}{c|}{0.001}      \\ \hline
\multicolumn{1}{l}{}    & \multicolumn{1}{l}{}           & \multicolumn{1}{l}{}            & \multicolumn{1}{l}{}           & \multicolumn{1}{l}{}            \\ \hline
\multicolumn{5}{|c|}{(c) Ho$_2$Ge$_2$O$_7$}                                                                                                                   \\ \hline
\multicolumn{1}{|c|}{}  & \multicolumn{1}{c|}{$E_{obs}$} & \multicolumn{1}{c|}{$E_{calc}$} & \multicolumn{1}{c|}{$I_{obs}$} & \multicolumn{1}{c|}{$I_{calc}$} \\ \hline
\multicolumn{1}{|c|}{D} & \multicolumn{1}{c|}{0}         & \multicolumn{1}{c|}{0}          & \multicolumn{1}{c|}{-}         & \multicolumn{1}{c|}{-}          \\ \hline
\multicolumn{1}{|c|}{S} & \multicolumn{1}{c|}{21.9(2)}   & \multicolumn{1}{c|}{21.8}       & \multicolumn{1}{c|}{0.05(2)}   & \multicolumn{1}{c|}{0.05}       \\ \hline
\multicolumn{1}{|c|}{D} & \multicolumn{1}{c|}{23.4(2)}   & \multicolumn{1}{c|}{23.6}       & \multicolumn{1}{c|}{0.22(5)}   & \multicolumn{1}{c|}{0.27}       \\ \hline
\multicolumn{1}{|c|}{D} & \multicolumn{1}{c|}{28.5(2)}   & \multicolumn{1}{c|}{28.4}       & \multicolumn{1}{c|}{0.19(2)}   & \multicolumn{1}{c|}{0.26}       \\ \hline
\multicolumn{1}{|c|}{S} & \multicolumn{1}{c|}{30.7(2)}   & \multicolumn{1}{c|}{30.7}       & \multicolumn{1}{c|}{0.05(2)}   & \multicolumn{1}{c|}{0.05}       \\ \hline
\multicolumn{1}{|c|}{D} & \multicolumn{1}{c|}{62.7(4)}   & \multicolumn{1}{c|}{63.2}       & \multicolumn{1}{c|}{1}         & \multicolumn{1}{c|}{1}          \\ \hline
\multicolumn{1}{|c|}{S} & \multicolumn{1}{c|}{75.5(5)}   & \multicolumn{1}{c|}{74.7}       & \multicolumn{1}{c|}{0.09(3)}   & \multicolumn{1}{c|}{0.05}       \\ \hline
\multicolumn{1}{|c|}{S} & \multicolumn{1}{c|}{79.0(4)}   & \multicolumn{1}{c|}{78.7}       & \multicolumn{1}{c|}{0.03(2)}   & \multicolumn{1}{c|}{0.03}       \\ \hline
\multicolumn{1}{|c|}{D} & \multicolumn{1}{c|}{-}         & \multicolumn{1}{c|}{81.4}       & \multicolumn{1}{c|}{-}         & \multicolumn{1}{c|}{0.007}      \\ \hline
\multicolumn{1}{|c|}{D} & \multicolumn{1}{c|}{87.0(4)}   & \multicolumn{1}{c|}{87.4}       & \multicolumn{1}{c|}{0.10(3)}   & \multicolumn{1}{c|}{0.06}       \\ \hline
\multicolumn{1}{|c|}{S} & \multicolumn{1}{c|}{-}         & \multicolumn{1}{c|}{91.0}       & \multicolumn{1}{c|}{-}         & \multicolumn{1}{c|}{0.001}      \\ \hline
\end{tabular}
\end{table}

\begin{table}[]
\centering
\caption{Composition of the vibronic bound state in (a) Ho$_2$Ti$_2$O$_7$, (b) Ho$_2$Sn$_2$O$_7$ and (c) Ho$_2$Ge$_2$O$_7$.}
\label{my-label}
\begin{tabular}{ccccccc}
\hline
\multicolumn{7}{|c|}{(a) Ho$_2$Ti$_2$O$_7$}                                                                                                                                                                                                                                              \\ \hline
\multicolumn{1}{|c|}{Energy (meV)} & \multicolumn{1}{c|}{$\ket{v1,0}$} & \multicolumn{1}{c|}{$\ket{v2,0}$} & \multicolumn{1}{c|}{$\ket{v_1,\hbar\omega_{0}}$} & \multicolumn{1}{c|}{$\ket{v_2,\hbar\omega_{0}}$} & \multicolumn{1}{c|}{$\ket{v3,0}$} & \multicolumn{1}{c|}{$\ket{v4,0}$} \\ \hline
\multicolumn{1}{|c|}{0}            & \multicolumn{1}{c|}{1}            & \multicolumn{1}{c|}{0}            & \multicolumn{1}{c|}{0}                           & \multicolumn{1}{c|}{0}                           & \multicolumn{1}{c|}{0}            & \multicolumn{1}{c|}{0}            \\ \hline
\multicolumn{1}{|c|}{0}            & \multicolumn{1}{c|}{0}            & \multicolumn{1}{c|}{1}            & \multicolumn{1}{c|}{0}                           & \multicolumn{1}{c|}{0}                           & \multicolumn{1}{c|}{0}            & \multicolumn{1}{c|}{0}            \\ \hline
\multicolumn{1}{|c|}{58.3}         & \multicolumn{1}{c|}{0}            & \multicolumn{1}{c|}{0}            & \multicolumn{1}{c|}{0}                           & \multicolumn{1}{c|}{-0.425}                      & \multicolumn{1}{c|}{0.905}        & \multicolumn{1}{c|}{0}            \\ \hline
\multicolumn{1}{|c|}{58.3}         & \multicolumn{1}{c|}{0}            & \multicolumn{1}{c|}{0}            & \multicolumn{1}{c|}{-0.425}                      & \multicolumn{1}{c|}{0}                           & \multicolumn{1}{c|}{0}            & \multicolumn{1}{c|}{-0.905}       \\ \hline
\multicolumn{1}{|c|}{61.8}         & \multicolumn{1}{c|}{0}            & \multicolumn{1}{c|}{0}            & \multicolumn{1}{c|}{0}                           & \multicolumn{1}{c|}{-0.905}                      & \multicolumn{1}{c|}{-0.425}       & \multicolumn{1}{c|}{0}            \\ \hline
\multicolumn{1}{|c|}{61.8}         & \multicolumn{1}{c|}{0}            & \multicolumn{1}{c|}{0}            & \multicolumn{1}{c|}{0.905}                       & \multicolumn{1}{c|}{0}                           & \multicolumn{1}{c|}{0}            & \multicolumn{1}{c|}{-0.425}       \\ \hline
\multicolumn{1}{l}{}               & \multicolumn{1}{l}{}              & \multicolumn{1}{l}{}              & \multicolumn{1}{l}{}                             & \multicolumn{1}{l}{}                             & \multicolumn{1}{l}{}              & \multicolumn{1}{l}{}              \\ \hline
\multicolumn{7}{|c|}{(b) Ho$_2$Sn$_2$O$_7$}                                                                                                                                                                                                                                              \\ \hline
\multicolumn{1}{|c|}{Energy (meV)} & \multicolumn{1}{c|}{$\ket{v1,0}$} & \multicolumn{1}{c|}{$\ket{v2,0}$} & \multicolumn{1}{c|}{$\ket{v_1,\hbar\omega_{0}}$} & \multicolumn{1}{c|}{$\ket{v_2,\hbar\omega_{0}}$} & \multicolumn{1}{c|}{$\ket{v3,0}$} & \multicolumn{1}{c|}{$\ket{v4,0}$} \\ \hline
\multicolumn{1}{|c|}{0}            & \multicolumn{1}{c|}{1}            & \multicolumn{1}{c|}{0}            & \multicolumn{1}{c|}{0}                           & \multicolumn{1}{c|}{0}                           & \multicolumn{1}{c|}{0}            & \multicolumn{1}{c|}{0}            \\ \hline
\multicolumn{1}{|c|}{0}            & \multicolumn{1}{c|}{0}            & \multicolumn{1}{c|}{1}            & \multicolumn{1}{c|}{0}                           & \multicolumn{1}{c|}{0}                           & \multicolumn{1}{c|}{0}            & \multicolumn{1}{c|}{0}            \\ \hline
\multicolumn{1}{|c|}{52.4}         & \multicolumn{1}{c|}{0}            & \multicolumn{1}{c|}{0}            & \multicolumn{1}{c|}{0}                           & \multicolumn{1}{c|}{-0.529}                      & \multicolumn{1}{c|}{0.848}        & \multicolumn{1}{c|}{0}            \\ \hline
\multicolumn{1}{|c|}{52.4}         & \multicolumn{1}{c|}{0}            & \multicolumn{1}{c|}{0}            & \multicolumn{1}{c|}{-0.529}                      & \multicolumn{1}{c|}{0}                           & \multicolumn{1}{c|}{0}            & \multicolumn{1}{c|}{-0.848}       \\ \hline
\multicolumn{1}{|c|}{55.0}         & \multicolumn{1}{c|}{0}            & \multicolumn{1}{c|}{0}            & \multicolumn{1}{c|}{0}                           & \multicolumn{1}{c|}{-0.848}                      & \multicolumn{1}{c|}{-0.529}       & \multicolumn{1}{c|}{0}            \\ \hline
\multicolumn{1}{|c|}{55.0}         & \multicolumn{1}{c|}{0}            & \multicolumn{1}{c|}{0}            & \multicolumn{1}{c|}{0.848}                       & \multicolumn{1}{c|}{0}                           & \multicolumn{1}{c|}{0}            & \multicolumn{1}{c|}{-0.529}       \\ \hline
\multicolumn{1}{l}{}               & \multicolumn{1}{l}{}              & \multicolumn{1}{l}{}              & \multicolumn{1}{l}{}                             & \multicolumn{1}{l}{}                             & \multicolumn{1}{l}{}              & \multicolumn{1}{l}{}              \\ \hline
\multicolumn{7}{|c|}{(c) Ho$_2$Ge$_2$O$_7$}                                                                                                                                                                                                                                              \\ \hline
\multicolumn{1}{|c|}{Energy (meV)} & \multicolumn{1}{c|}{$\ket{v1,0}$} & \multicolumn{1}{c|}{$\ket{v2,0}$} & \multicolumn{1}{c|}{$\ket{v_1,\hbar\omega_{0}}$} & \multicolumn{1}{c|}{$\ket{v_2,\hbar\omega_{0}}$} & \multicolumn{1}{c|}{$\ket{v3,0}$} & \multicolumn{1}{c|}{$\ket{v4,0}$} \\ \hline
\multicolumn{1}{|c|}{0}            & \multicolumn{1}{c|}{1}            & \multicolumn{1}{c|}{0}            & \multicolumn{1}{c|}{0}                           & \multicolumn{1}{c|}{0}                           & \multicolumn{1}{c|}{0}            & \multicolumn{1}{c|}{0}            \\ \hline
\multicolumn{1}{|c|}{0}            & \multicolumn{1}{c|}{0}            & \multicolumn{1}{c|}{1}            & \multicolumn{1}{c|}{0}                           & \multicolumn{1}{c|}{0}                           & \multicolumn{1}{c|}{0}            & \multicolumn{1}{c|}{0}            \\ \hline
\multicolumn{1}{|c|}{60.6}         & \multicolumn{1}{c|}{0}            & \multicolumn{1}{c|}{0}            & \multicolumn{1}{c|}{0}                           & \multicolumn{1}{c|}{-0.494}                      & \multicolumn{1}{c|}{0.870}        & \multicolumn{1}{c|}{0}            \\ \hline
\multicolumn{1}{|c|}{60.6}         & \multicolumn{1}{c|}{0}            & \multicolumn{1}{c|}{0}            & \multicolumn{1}{c|}{-0.494}                      & \multicolumn{1}{c|}{0}                           & \multicolumn{1}{c|}{0}            & \multicolumn{1}{c|}{-0.870}       \\ \hline
\multicolumn{1}{|c|}{63.4}         & \multicolumn{1}{c|}{0}            & \multicolumn{1}{c|}{0}            & \multicolumn{1}{c|}{0}                           & \multicolumn{1}{c|}{-0.870}                      & \multicolumn{1}{c|}{-0.494}       & \multicolumn{1}{c|}{0}            \\ \hline
\multicolumn{1}{|c|}{63.4}         & \multicolumn{1}{c|}{0}            & \multicolumn{1}{c|}{0}            & \multicolumn{1}{c|}{0.870}                       & \multicolumn{1}{c|}{0}                           & \multicolumn{1}{c|}{0}            & \multicolumn{1}{c|}{-0.494}       \\ \hline
\end{tabular}
\end{table}

\bibliography{Ho_CEF_bib}